\documentclass[aps, twocolumn, prl, superscriptaddress, nofootinbib]{revtex4-2}

\pdfoutput=1
\usepackage{graphicx}
\usepackage{amsmath}
\usepackage{dcolumn}
\usepackage{nicefrac}
\usepackage[T1]{fontenc}
\usepackage[activate={true,nocompatibility}]{microtype}
\usepackage[colorlinks]{hyperref}
\hypersetup{
    colorlinks=true,
    linkcolor=black,
    filecolor=black,
    citecolor=black,
    urlcolor=blue
  }
  
\newcommand{\isotope}[4]{\mbox{\ensuremath{^{#1}_{#2}\textrm{#3}_{#4}}}}
\newcommand{\isot}[2]{\ensuremath{^{#1}\textrm{#2}}}

\newcommand{\Thalf}{\ensuremath{T_{\nicefrac{1}{2}}} }

\begin{document}

\title{Microsecond Isomer at the N=20 Island of Shape Inversion Observed at FRIB}

\author{T.~J.~Gray}
\affiliation{Physics Division, Oak Ridge National Laboratory, Oak Ridge Tennessee 37831, USA}
\author{J.~M.~Allmond}
\affiliation{Physics Division, Oak Ridge National Laboratory, Oak Ridge Tennessee 37831, USA}

\author{Z.~Xu}
\affiliation{Department of Physics and Astronomy, University of Tennessee, Knoxville, Tennessee, 37966, USA}
\author{T.~T.~King}
\affiliation{Physics Division, Oak Ridge National Laboratory, Oak Ridge Tennessee 37831, USA}
\author{R.~S.~Lubna}
\affiliation{Facility for Rare Isotope Beams, Michigan State University, East Lansing, Michigan, 48824, USA}

\author{H.~L.~Crawford}
\affiliation{Nuclear Science Division, Lawrence Berkeley National Laboratory, Berkeley, California, 94720, USA}
\author{V.~Tripathi}
\affiliation{Department of Physics, Florida State University, Tallahassee, Florida, 32306, USA}
\author{B.~P.~Crider}
\affiliation{Department of Physics and Astronomy, Mississippi State University, Mississippi State, Mississippi 39762, USA}
\author{R.~Grzywacz}
\affiliation{Department of Physics and Astronomy, University of Tennessee, Knoxville, Tennessee, 37966, USA}
\affiliation{Physics Division, Oak Ridge National Laboratory, Oak Ridge Tennessee 37831, USA}
\author{S.~N.~Liddick}
\affiliation{Facility for Rare Isotope Beams, Michigan State University, East Lansing, Michigan, 48824, USA}
\affiliation{Department of Chemistry, Michigan State University, East Lansing, Michigan, 48824, USA}

\author{A.~O.~Macchiavelli}
\affiliation{Physics Division, Oak Ridge National Laboratory, Oak Ridge Tennessee 37831, USA}
\author{T.~Miyagi}
\affiliation{Department of Physics, Technische Universit\"{a}t Darmstadt, Darmstadt, Germany}
\affiliation{ExtreMe Matter Institute EMMI, GSI Helmholtzzentrum f\"ur Schwerionenforschung GmbH, 64291 Darmstadt, Germany}
\author{A.~Poves}
\affiliation{Departamento de F\`{i}sica Te\'{o}rica and IFT-UAM/CSIC, Universidad Aut\'{o}noma de Madrid, E-28049 Madrid, Spain}

\author{A.~Andalib}
\affiliation{Facility for Rare Isotope Beams, Michigan State University, East Lansing, Michigan, 48824, USA}
\affiliation{Department of Physics and Astronomy, Michigan State University, East Lansing, Michigan, 48824, USA}
\author{E.~Argo}
\affiliation{Facility for Rare Isotope Beams, Michigan State University, East Lansing, Michigan, 48824, USA}
\affiliation{Department of Physics and Astronomy, Michigan State University, East Lansing, Michigan, 48824, USA}
\author{C.~Benetti}
\affiliation{Department of Physics, Florida State University, Tallahassee, Florida, 32306, USA}
\author{S.~Bhattacharya}
\affiliation{Department of Physics, Florida State University, Tallahassee, Florida, 32306, USA}
\author{C.~M.~Campbell}
\affiliation{Nuclear Science Division, Lawrence Berkeley National Laboratory, Berkeley, California, 94720, USA}
\author{M.~P.~Carpenter}
\affiliation{Argonne National Laboratory, Argonne, Illinois 60439, USA}
\author{J.~Chan}
\affiliation{Department of Physics and Astronomy, University of Tennessee, Knoxville, Tennessee, 37966, USA}
\author{A.~Chester}
\affiliation{Facility for Rare Isotope Beams, Michigan State University, East Lansing, Michigan, 48824, USA}
\author{J.~Christie}
\affiliation{Department of Physics and Astronomy, University of Tennessee, Knoxville, Tennessee, 37966, USA}
\author{B.~R.~Clark}
\affiliation{Department of Physics and Astronomy, Mississippi State University, Mississippi State, Mississippi 39762, USA}
\author{I.~Cox}
\affiliation{Department of Physics and Astronomy, University of Tennessee, Knoxville, Tennessee, 37966, USA}
\author{A.~A.~Doetsch}
\affiliation{Facility for Rare Isotope Beams, Michigan State University, East Lansing, Michigan, 48824, USA}
\affiliation{Department of Physics and Astronomy, Michigan State University, East Lansing, Michigan, 48824, USA}
\author{J.~Dopfer}
\affiliation{Facility for Rare Isotope Beams, Michigan State University, East Lansing, Michigan, 48824, USA}
\affiliation{Department of Physics and Astronomy, Michigan State University, East Lansing, Michigan, 48824, USA}
\author{J.~G.~Duarte}
\affiliation{Lawrence Livermore National Laboratory, Livermore, California 94550, USA}
\author{P.~Fallon}
\affiliation{Nuclear Science Division, Lawrence Berkeley National Laboratory, Berkeley, California, 94720, USA}
\author{A.~Frotscher}
\affiliation{Nuclear Science Division, Lawrence Berkeley National Laboratory, Berkeley, California, 94720, USA}
\author{T.~Gaballah}
\affiliation{Department of Physics and Astronomy, Mississippi State University, Mississippi State, Mississippi 39762, USA}
\author{J.~T.~Harke}
\affiliation{Lawrence Livermore National Laboratory, Livermore, California 94550, USA}
\author{J.~Heideman}
\affiliation{Department of Physics and Astronomy, University of Tennessee, Knoxville, Tennessee, 37966, USA}
\author{H.~Huegen}
\affiliation{Department of Physics and Astronomy, University of Tennessee, Knoxville, Tennessee, 37966, USA}
\author{J. D. Holt}
\affiliation{TRIUMF, 4004 Wesbrook Mall, Vancouver, BC V6T 2A3, Canada}
\affiliation{Department of Physics, McGill University, Montr\'eal, QC H3A 2T8, Canada}
\author{R.~Jain}
\affiliation{Facility for Rare Isotope Beams, Michigan State University, East Lansing, Michigan, 48824, USA}
\affiliation{Department of Physics and Astronomy, Michigan State University, East Lansing, Michigan, 48824, USA}
\author{N.~Kitamura}
\affiliation{Department of Physics and Astronomy, University of Tennessee, Knoxville, Tennessee, 37966, USA}
\author{K.~Kolos}
\affiliation{Lawrence Livermore National Laboratory, Livermore, California 94550, USA}
\author{F.~G.~Kondev}
\affiliation{Argonne National Laboratory, Argonne, Illinois 60439, USA}
\author{A.~Laminack}
\affiliation{Physics Division, Oak Ridge National Laboratory, Oak Ridge Tennessee 37831, USA}
\author{B.~Longfellow}
\affiliation{Lawrence Livermore National Laboratory, Livermore, California 94550, USA}
\author{S.~Luitel}
\affiliation{Department of Physics and Astronomy, Mississippi State University, Mississippi State, Mississippi 39762, USA}
\author{M.~Madurga}
\affiliation{Department of Physics and Astronomy, University of Tennessee, Knoxville, Tennessee, 37966, USA}
\author{R.~Mahajan}
\affiliation{Facility for Rare Isotope Beams, Michigan State University, East Lansing, Michigan, 48824, USA}
\author{M.~J.~Mogannam}
\affiliation{Facility for Rare Isotope Beams, Michigan State University, East Lansing, Michigan, 48824, USA}
\affiliation{Department of Chemistry, Michigan State University, East Lansing, Michigan, 48824, USA}
\author{C.~Morse}
\affiliation{National Nuclear Data Center, Brookhaven National Laboratory, Upton, New York 11973, USA}
\author{S.~Neupane}
\affiliation{Department of Physics and Astronomy, University of Tennessee, Knoxville, Tennessee, 37966, USA}
\author{A.~Nowicki}
\affiliation{Department of Physics and Astronomy, University of Tennessee, Knoxville, Tennessee, 37966, USA}
\author{T.~H.~Ogunbeku}
\affiliation{Department of Physics and Astronomy, Mississippi State University, Mississippi State, Mississippi 39762, USA}
\affiliation{Facility for Rare Isotope Beams, Michigan State University, East Lansing, Michigan, 48824, USA}
\author{W.-J.~Ong}
\affiliation{Lawrence Livermore National Laboratory, Livermore, California 94550, USA}
\author{C.~Porzio}
\affiliation{Nuclear Science Division, Lawrence Berkeley National Laboratory, Berkeley, California, 94720, USA}
\author{C.~J.~Prokop}
\affiliation{Los Alamos National Laboratory, Los Alamos, New Mexico 87545, USA}
\author{B.~C.~Rasco}
\affiliation{Physics Division, Oak Ridge National Laboratory, Oak Ridge Tennessee 37831, USA}
\author{E.~K.~Ronning}
\affiliation{Facility for Rare Isotope Beams, Michigan State University, East Lansing, Michigan, 48824, USA}
\affiliation{Department of Chemistry, Michigan State University, East Lansing, Michigan, 48824, USA}
\author{E.~Rubino}
\affiliation{Facility for Rare Isotope Beams, Michigan State University, East Lansing, Michigan, 48824, USA}
\author{T.~J.~Ruland}
\affiliation{Department of Physics and Astronomy, Louisiana State University, Baton Rouge, Louisiana 70803, USA}
\author{K.~P.~Rykaczewski}
\affiliation{Physics Division, Oak Ridge National Laboratory, Oak Ridge Tennessee 37831, USA}
\author{L.~Schaedig}
\affiliation{Facility for Rare Isotope Beams, Michigan State University, East Lansing, Michigan, 48824, USA}
\affiliation{Department of Physics and Astronomy, Michigan State University, East Lansing, Michigan, 48824, USA}
\author{D.~Seweryniak}
\affiliation{Argonne National Laboratory, Argonne, Illinois 60439, USA}
\author{K.~Siegl}
\affiliation{Department of Physics and Astronomy, University of Tennessee, Knoxville, Tennessee, 37966, USA}
\author{M.~Singh}
\affiliation{Department of Physics and Astronomy, University of Tennessee, Knoxville, Tennessee, 37966, USA}
\author{A.~E.~Stuchbery}
\affiliation{Department of Nuclear Physics and Accelerator Applications, Research School of Physics, Australian National University, Canberra, ACT, 2601, Australia}
\author{S.~L.~Tabor}
\affiliation{Department of Physics, Florida State University, Tallahassee, Florida, 32306, USA}
\author{T.~L.~Tang}
\affiliation{Department of Physics, Florida State University, Tallahassee, Florida, 32306, USA}
\author{T.~Wheeler}
\affiliation{Facility for Rare Isotope Beams, Michigan State University, East Lansing, Michigan, 48824, USA}
\affiliation{Department of Physics and Astronomy, Michigan State University, East Lansing, Michigan, 48824, USA}
\author{J.~A.~Winger}
\affiliation{Department of Physics and Astronomy, Mississippi State University, Mississippi State, Mississippi 39762, USA}
\author{J.~L.~Wood}
\affiliation{School of Physics, Georgia Institute of Technology, Atlanta, Georgia, 30332-0430, USA}

\begin{abstract}
Excited-state spectroscopy from the first Facility for Rare Isotope Beams (FRIB) experiment is reported. A 24(2)-$\mu$s isomer was observed with the FRIB Decay Station initiator (FDSi) through a cascade of 224- and 401-keV $\gamma$ rays in coincidence with $^{32}\textrm{Na}$ nuclei. This is the only known microsecond isomer ($1{\text{ }\mu\text{s}}\leq T_{1/2} < 1\text{ ms}$) in the region. This nucleus is at the heart of the $N=20$ island of shape inversion and is at the crossroads of spherical shell-model, deformed shell-model, and \textit{ab initio} theories. It can be represented as the coupling of a proton hole and neutron particle to $^{32}\textrm{Mg}$, $^{32}\textrm{Mg}+\pi^{-1} + \nu^{+1}$. This odd-odd coupling and isomer formation provides a sensitive measure of the underlying shape degrees of freedom of $^{32}\textrm{Mg}$, where the onset of spherical-to-deformed shape inversion begins with a low-lying deformed $2^+$ state at 885~keV and a low-lying shape-coexisting $0_2^+$ state at 1058~keV. We suggest two possible explanations for the 625-keV isomer in $^{32}$Na: a $6^-$ spherical shape isomer that decays by $E2$ or a $0^+$ deformed spin isomer that decays by $M2$. The present results and calculations are most consistent with the latter, indicating that the low-lying states are dominated by deformation.


\end{abstract}


\maketitle


\paragraph*{}%
The atomic nucleus is a self-organizing finite quantum many-body system. Among the phenomena exhibited by this system is deformation. Most nuclei are quadrupole deformed but spherical nuclei can be found along proton and/or neutron closed (magic) shells, e.g., 2, 8, 20, 28, 50, 82, 126. In addition, many nuclei manifest shape coexistence between the ground and excited states, e.g., from multi-particle, multi-hole (mp-nh) cross-shell excitations with orbital-dependent pairing and quadrupole-deformation correlation energies \cite{Heyde1983,Wood1992,Heyde2011}. Shape coexistence of a spherical or deformed ground state with a deformed excited state has been observed. However, shape coexistence where the ground state is deformed and the excited state is spherical has been more elusive.


\paragraph*{}
An example of shape coexistence is seen in the semi-magic $N=20$ isotonic chain. While doubly magic \isotope{40}{20}{Ca}{20} has a spherical $0_1^+$ ground state, excited $0^+$ states with normal and super deformation exist at 3.4 and 5.2~MeV, respectively~\cite{Wood1992,Ideguchi2001,Ideguchi2022}. Moving to semi-magic \isotope{38}{18}{Ar}{20}, \isotope{36}{16}{S}{20}, and \isotope{34}{14}{Si}{20}, the excited $0_2^+$ states progress from 3.4 to 2.7 MeV; the first-excited $2_1^+$ states vary from 2.2 to 3.3~MeV~\cite{NDS34,NDS36,NDS38}, consistent with spherical ground states. Then at \isotope{32}{12}{Mg}{20}, a low first-excited $2_1^+$ energy of 885~keV is observed with a $0_2^+$ state at 1058~keV, consistent with a deformed ground state --- despite the $N=20$ closed neutron shell. In fact, the spacing of the 0-2-4-6 rotational sequence of the \isot{32}{Mg} ground band (885, 2322, 4095~keV) is similar to the excited band (560, 1973, 3911~keV) in \isot{38}{Ar}, starting with the 3.4~MeV $0_2^{+}$ state. This crossing is known as the ``island of inversion'', where the deformed ``intruder'' configuration drops below the spherical one. See Refs.~\cite{Thibault1975,Campi1975,Huber1978,Detrz1979,GuillemaudMueller1984,Doornebal2009,Sorlin2008} and Figs.~3, 44 and 45 of Ref.~\cite{Heyde2011}. 

\paragraph*{} 
 Shape inversion is now being recognized as a prominent phenomenon for neutron-rich nuclei beyond the $N=20$ region. An open question remains as to whether the excited spherical states are ever preserved after the inversion and in the presence of a deformed ground state, where large configuration mixing may dominate~\cite{Wimmer2010,Fortune2011,Crawford2016,Lay2014,Macchiavelli2017,Elder2019,Caurier2004}, particularly near the crossing. For instance, it was proposed in Refs.~\cite{Crawford2016,Macchiavelli2017} that strong mixing of three $0^+$ configurations (two deformed, 2p-2h and 4p-4h, and one spherical, 0p-0h) was required to explain (t,p)\isot{32}{Mg} data \cite{Wimmer2010}, implying an overall structure dominated by deformation. Ref.~\cite{Elder2019} measured the lifetime and population of the $0_2^{+}$ state in a $^{34}$Si two-proton knockout reaction, finding a large $B(E2; 2_1^{+} \rightarrow 0_2^{+})$, and low population of the $0_2^{+}$ state, indicating a reduction of 0p-0h character. Interestingly, three-state mixing was also required to explain the weak $0^+_3\rightarrow0_1^+$ $E0$ decay of \isot{40}{Ca}~\cite{Ideguchi2022}.


\paragraph*{}
Long-lived excited states (isomers) can be used as sensitive probes of nuclear structures~\cite{Grzywacz1995,Grzywacz1998}, due in part to the limited number of configuration combinations that can result in isomerism. Isomers in the odd-proton \isot{32,34}{Al} isotopes, which sit between \textrm{Si} and \textrm{Mg}, have been used to probe the transitional region between regular and inverted structures. While the $J^\pi = 4^+$, $\Thalf = 200(20)$~ns isomer in \isot{32}{Al} and its ground state can be explained by purely spherical structures~\cite{Robinson2016,Grevy2004,Fornal1997}, the $J^\pi = 1^+$, $\Thalf = 26(1)$~ms isomer in \isot{34}{Al} seems to be a deformed $\nu$2p-1h excitation with a ground state containing a 50/50 admixture of spherical and deformed configurations~\cite{Pritychenko2001, Himpe2008, Nociforo2012,Rotaru2012, Lica2017, Xu2018}. In the transitional region of the $N=28$ island of inversion, \isot{43}{S} has a $J^\pi = 7/2^-$, $\Thalf = 415(5)$~ns isomer, initially thought to be spherical~\cite{Sarazin2000,Gaudefroy2009}, in coexistence with a deformed ground state. However, the isomer is now understood to be a weakly deformed rotational band head~\cite{Longfellow2020}, consistent with the suggestion that the configurations are strongly mixed in the region of the crossing~\cite{Caurier2004}. These isomers have played decisive roles in the structural interpretation of these transitional regions.


\paragraph*{}
Here we examine the odd-odd nucleus \isot{32}{Na} ($Z=11$, $N=21$), which is one-proton hole and one-neutron particle outside of \isot{32}{Mg} --- this is beyond the transitional region and firmly within the island of inversion, where excited spherical states may be preserved due to the larger energy spacing between deformed and spherical configurations. We report a new micro-second isomer in \isot{32}{Na} with two potential explanations: one relating to a spherical shape isomer that decays by $E2$ $\gamma$-ray emission, and the other a deformed spin isomer that decays by $M2$ $\gamma$-ray emission.



\paragraph*{}
The present work follows from the first Facility for Rare Isotope Beams (FRIB) experiment~\cite{Crawford2022}, with the analysis expanded to excited-state spectroscopy. Details of the experimental setup can be found in Ref.~\cite{Crawford2022}. In short, a primary beam of 172.3-MeV/u \isot{48}{Ca} at a power of $\sim$1~kW was incident on a \isot{9}{Be} primary target. The Advanced Rare Isotope Separator (ARIS)~\cite{ARIS} selected a cocktail beam, focused on \isot{42}{Si}, in which \isot{32}{Na} was present. This cocktail beam was delivered to the FRIB Decay Station initiator (FDSi)~\cite{FDSi1,FDSi2}, and each nucleus was identified on an event-by-event basis by the energy loss through an upstream Si detector and the time-of-flight from the ARIS separator. See Fig.~1 of Ref.~\cite{Crawford2022} for a complete particle identification plot, which roughly spans $N=20-28$. 

\paragraph*{}%
 An implantation detector made of yttrium orthosilicate (Y$_2$SiO$_5$, YSO) ~\cite{IMPLANT} was positioned at the center of the FDSi discrete focal point. On the north side of the beam line, a $\gamma$-ray detector array, DEcay Germanium Array initiator (DEGAi), consisting of 11 HPGe clover detectors and 15 fast-timing LaBr$_3$ detectors, was present. On the south side, the 88 modules of the NEutron Xn Tracking array initiator (NEXTi) were arranged in a double-layer arch, which is an expansion of the VANDLE array to measure neutron time-of-flight with one-meter flight paths~\cite{VANDLE1, VANDLE2}. The results presented here focus on the YSO implant and HPGe DEGAi array, namely $\gamma$-ray energies, and time differences between implanted ions and subsequent $\gamma$ rays.

\begin{figure}[t]
\includegraphics[width=\columnwidth]{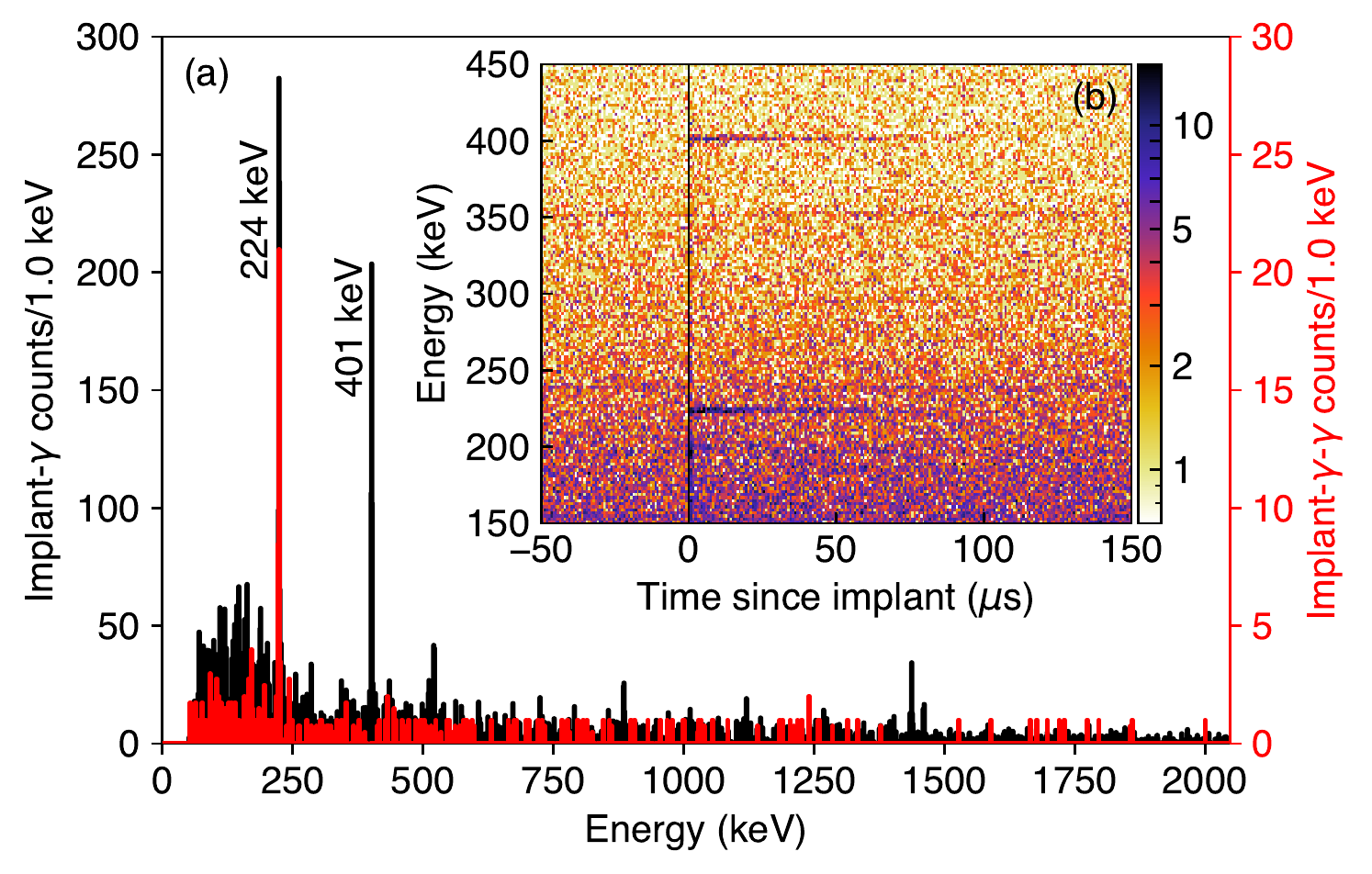}
\caption{(a): $\gamma$-ray spectrum following implanted \isot{32}{Na} ions (black), and $\gamma$-ray spectrum in coincidence with 401-keV $\gamma$ rays and following implanted \isot{32}{Na} ions (red). (b): $\gamma$-ray energy versus time difference of $\gamma$ ray and \isot{32}{Na} implant. }\label{fig:na32_energy}
\end{figure}

\paragraph*{}%
The $\gamma$ rays following implanted \isot{32}{Na} ions and the time distribution between them, which expands to 150~$\mu$s, are shown in Fig.~\ref{fig:na32_energy}. Two $\gamma$-ray peaks with equal efficiency-corrected intensity are seen at 224 and 401~keV, cf.~Fig.~\ref{fig:na32_energy}(a)(black). Further, these two peaks are in coincidence with each other, cf.~Fig.~\ref{fig:na32_energy}(a)(red), and they show exponential decay from the prompt x-ray/$\gamma$-ray flash, which is shown as the black vertical line in Fig.~\ref{fig:na32_energy}(b), induced by ion implantation at $t=0$. Several room background $\gamma$ rays are also present before and after this prompt flash, i.e. 239~keV from \isot{212}{Bi} and 352~keV from \isot{214}{Bi}. The data from negative $\gamma$-implant times can be used for background subtraction.


\begin{figure}[t]
    \includegraphics[width=\columnwidth]{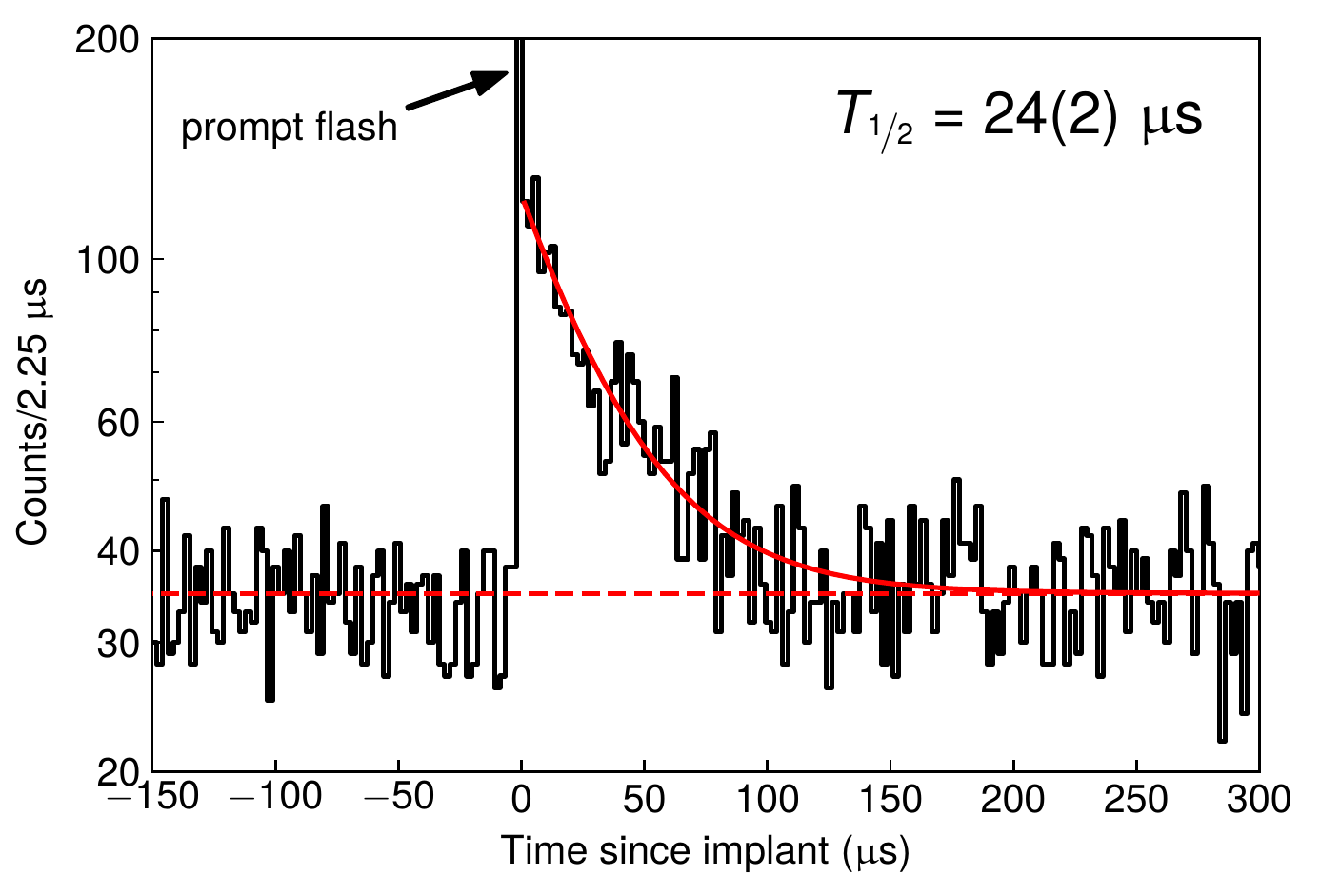}
    \caption{The $\gamma$-implant time distribution formed by the sum of 224- and 401-keV $\gamma$ gates. An exponential maximum-likelihood fit is shown with the solid red line. The dashed red line shows the constant baseline. }\label{fig:na32_time}
\end{figure}

\paragraph*{}%
An isomeric half-life of $\Thalf=24(2)$~$\mu$s is determined from a maximum-likelihood fit to the $\gamma$-implant time distribution, cf.~Fig.~\ref{fig:na32_time}, where gates on the 224- and \mbox{401-keV} $\gamma$ rays were combined. The isomer constituted $1.8(3)\%$ of the total \isot{32}{Na} beam, where the momentum acceptance for \isot{32}{Na} was $1\%$ and centered at $1.6\%$ higher than the predicted optimum momentum; the relatively low yield could be indicative of a non-yrast structure assignment~\cite{Grzywacz1995,Daugas2001}. Ultimately, the results indicate a microsecond isomer in \isot{32}{Na} which decays by a cascade of two $\gamma$ rays to either the ground state or a long-lived $\beta$-decaying state in \isot{32}{Na}. More details on the two possibilities are given in the calculations below. Alternatively, there could be an additional $\gamma$ ray that is unobserved due to the present detection limit of $\approx 70$~keV. No other microsecond ``beam'' isomers were observed in the data for nuclei spanning $N=20-28$, cf.~Fig.~1 of Ref.~\cite{Crawford2022}; this is the only microsecond isomer ($1{\text{ }\mu\text{s}}\leq \Thalf < 1\text{ ms}$) currently known in the region.



\paragraph*{}
The multipolarity of the depopulating transition can be inferred using the measured $\gamma$-ray energies and lifetime. $E1$ and $M1$ transitions would require unreasonably hindered transition strengths, e.g., $B(M1{\downarrow}) \lesssim 10^{-7}$~W.u., to obtain the observed $24(2)$~$\mu$s. Likewise, $E3$ and $M3$ transitions would require unreasonably enhanced transition strengths, e.g., $B(E3{\downarrow}) \gtrsim 500$~W.u.~which is a factor of 10 larger than the recommended upper limit for $5\leq A \leq 44$ \cite{Endt1993,TOI}. This leaves an $E2$ or $M2$ depopulating transition as the most likely scenario with the following possibilities: $B(E2{\downarrow};\, 224 \text{ keV}) = 0.0069(6)$~W.u., $B(E2{\downarrow};\, 401 \text{ keV}) = 0.00038(3)$~W.u., $B(M2{\downarrow};\, 224 \text{ keV}) = 0.23(2)$~W.u., or $B(M2{\downarrow};\, 401 \text{ keV}) = 0.012(1)$~W.u.. We tentatively assign the 224-keV $\gamma$ ray as the depopulating transition of the isomer, as this requires lower hindrance factors.



\begin{figure}[t]
 \includegraphics[width=\columnwidth]{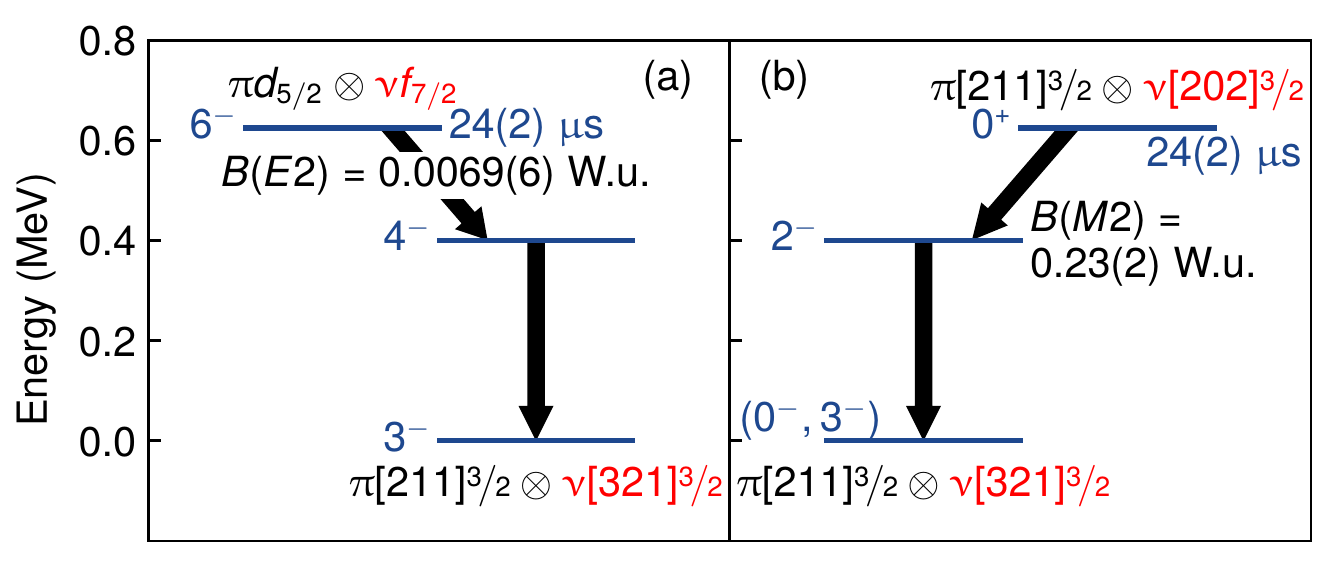}
 \caption{Schematic representation of (a) $6^-$ spherical shape isomer and (b) $0^+$ deformed spin isomer.}
 \label{fig:schematic}
\end{figure}

\paragraph*{}%
For low-lying configurations within the spherical shell model, the odd proton occupies the $\pi d_{5/2}$ orbital, while the odd neutron occupies the $\nu f_{7/2}$ orbital. The coupling of these give a multiplet with $J^{\pi} = 1^{-}...6^{-}$. The most simple expectation is that the $1^-$ and $6^-$ multiplet members would be lowest in energy~\cite{Paar1979}. However, multi-particle interactions could perturb the expected energy ordering of the multiplet states. For low-lying configurations within the deformed shell model (Nilsson model)~\cite{Nilsson1955,ShapesAndShells} at a deformation of $\epsilon_2 \approx +0.4$ (adopted from fitting the energies and $E2$ strengths of neighboring nuclei), the odd proton is expected to occupy the $\pi [211]3/2^{+}$ orbital, while the odd neutron should occupy the $\nu [321]3/2^-$ orbital. The coupling of these two orbitals gives rise to two low-lying states: one with $J^\pi = 3^{-}$ (parallel coupling), and a second with $J^\pi = 0^-$ (anti-parallel coupling). Both of these will have rotational states built upon them. In addition, the neutron could be excited to the nearby $\nu [202]3/2^+$ orbital, giving rise to a low-lying $J^\pi = 0^+$ state. Based on these expected configurations, two isomeric scenarios can be formed involving $E2$ and $M2$ transitions, cf.~Fig.~\ref{fig:schematic}.

\paragraph*{}%

\paragraph*{}%
A $6^-$ spherical shape isomer --- based on the $\pi d_{5/2}\otimes\nu f_{7/2}$ multiplet --- would decay via a hindered $E2$ transition to the deformed $4^-$ member of the $K=3^-$ rotational band. Thus, the spin sequence is $6^{-} \rightarrow 4^{-} \rightarrow 3^{-}$, giving the two observed $\gamma$ rays. The observed weak transition strength of $B(E2{\downarrow})=0.0069(6)$~W.u.~is consistent with shape coexistence: the spherical $6^-$ state may only decay to the deformed $4^-$ state by weak mixing between spherical and deformed $6^-$ and $4^-$ configurations. A range of two-state mixing solutions reproduce the observed transition strength, requiring mixing angles of $\theta_6 < 2^\circ$ and $\theta_4 < 10^\circ$ for the $6^-$ and $4^-$ states, respectively. This scenario would constitute shape coexistence between a deformed ground state and spherical excited state with minimal mixing of the configurations.



\paragraph*{}%
A $0^+$ deformed spin isomer --- based on the $\pi [211]3/2^{+} \otimes \nu [202]3/2^+$, $J^{\pi} = 0^+$ deformed band-head --- would decay via an $M2$ transition, with a non quenched $B(M2{\downarrow})\approx 1$~W.u., to the $2^-$ rotational member of the $K=0^{-}$ band. Thus, the experimental value of $B(M2{\downarrow}) = 0.23(2)$~W.u. indicates a hindrance factor of 5 for this scenario. The spin sequence in this case is $0^{+} \rightarrow 2^{-} \rightarrow 0^{-}$. Alternatively, if the $\pi [211]3/2^{+} \otimes \nu [321]3/2^-$, $J^\pi = 3^-$ state is near or below the $0^-$ state, instead of near or above the $2^-$ state, we would observe the $0^{+} \rightarrow 2^{-} \rightarrow 3^{-}$ sequence instead. This $0^+$ isomer scenario would imply that the low-lying states are dominated by deformation with no clear remnants of the spherical shape.

\begin{figure}[t]
\includegraphics[width=\columnwidth]{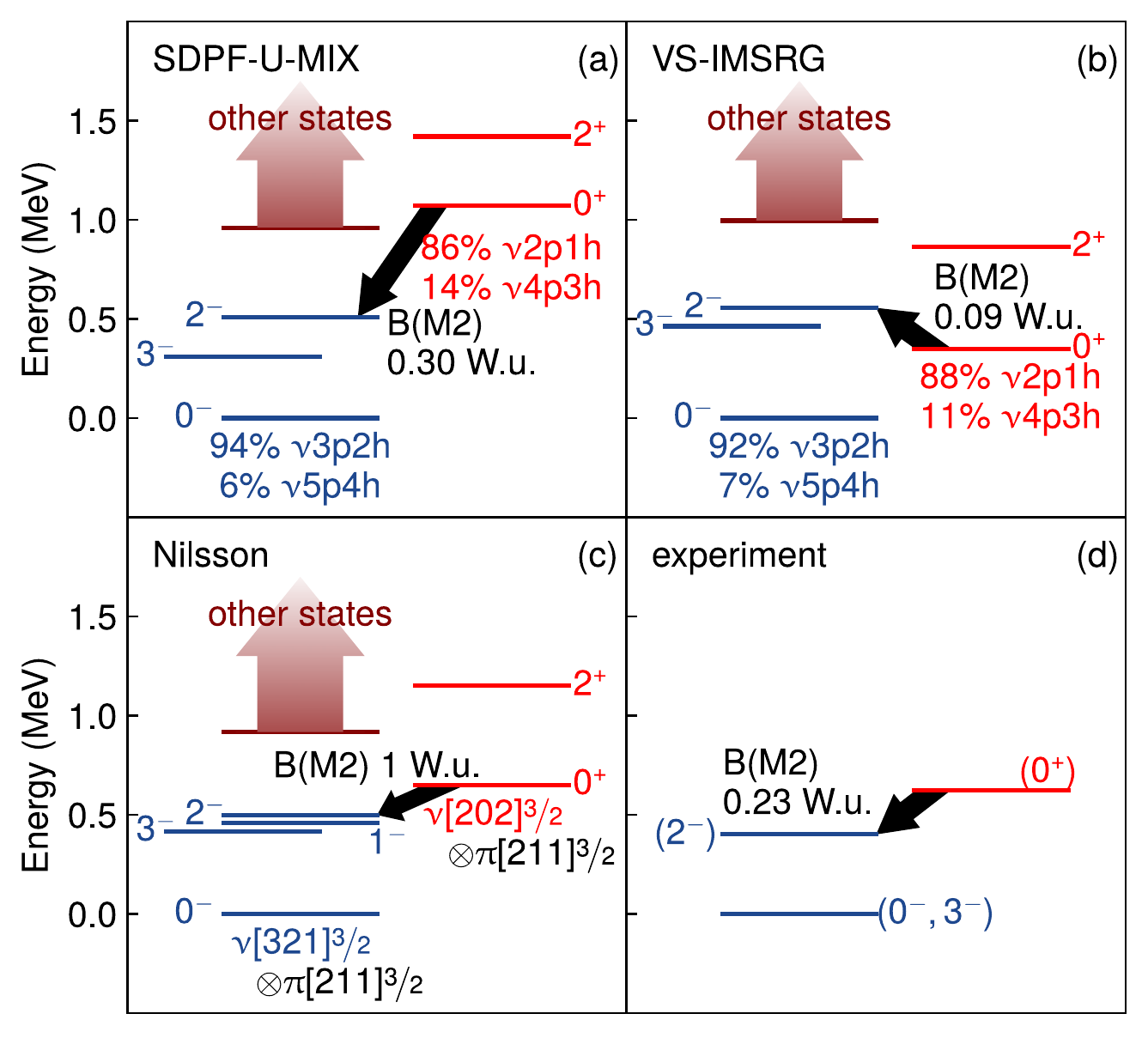}
\caption{Comparison of calculated and proposed experimental level schemes. Deformed $0^{-}$, $3^{-}$, and $0^{+}$ band-heads are present with associated rotational bands. The ``mp-nh'' language refers to the number of neutron particles above and neutron holes below the $N=20$ shell closure, where mixing of the cross-shell excitations is predicted.}
\label{fig:calc_summary}
\end{figure}

\begin{figure}[t]
\includegraphics[width=\columnwidth]{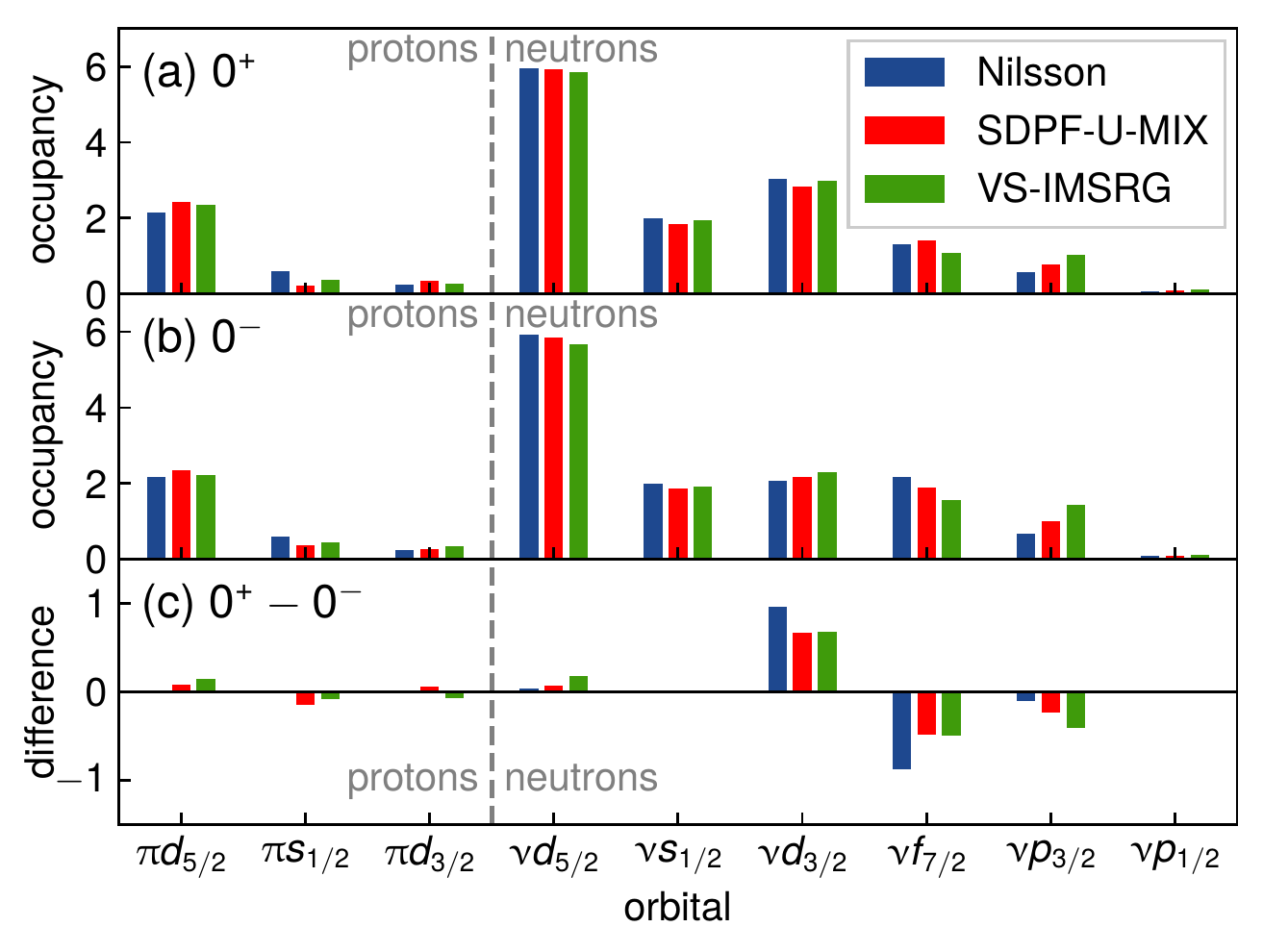}
\caption{Single-particle occupancies from Nilsson, SDPF-U-MIX, and VS-IMSRG calculations: (a) Occupancies for the $0^{+}$ state, (b): Occupancies for the $0^{-}$ state, and (c) Difference between occupancies of the $0^{+}$ and $0^{-}$ states, highlighting the strong $\nu f_{7/2}\rightarrow \nu d_{3/2}$ nature of the $M2$ transition. The occupancies of the $2^{-}$ and $0^{-}$ states are near identical in all three calculations, where the former is a rotation built upon the latter.}
\label{fig:occupancies}
\end{figure}

\paragraph*{}
Several state-of-the-art calculations were run to further access the isomer and experimental decay scheme. The results are given in Fig.~\ref{fig:calc_summary}: (a) Large-scale shell-model calculations ``SDPF-U-MIX'' in the $\pi (s_{1/2}, d_{5/2}, d_{3/2}) \nu (s_{1/2}, d_{5/2}, d_{3/2}, f_{7/2}, p_{3/2}, p_{1/2}, f_{5/2})$ space with effective charges $e_n = 0.46$ and $e_p = 1.31$~\cite{Caurier2014,Nowaki2009, Poves2014}. The shell-model calculations used the ANTOINE code~\cite{ANTOINE} for diagonalization, allowing a maximum of 6 neutrons above $N=20$. (b) \textit{Ab initio} calculations ``VS-IMSRG'' using the 1.8/2.0 (EM) interaction~\cite{Hebeler2011} with the imsrg++~\cite{imsrg} code (see Refs.~\cite{Miyagi2022, Simonis2017, Stroberg2021, Stroberg2017, Stroberg2019, Miyagi2020}). The diagonalization was carried out using KSHELL~\cite{Shimizu2019}, in the $\pi (s_{1/2}, d_{5/2}, d_{3/2}) \nu (s_{1/2}, d_{5/2}, d_{3/2}, f_{7/2}, p_{3/2}, p_{1/2})$ space with no truncation. Both ``SDPF-U-MIX'' and ``VS-IMSRG'' calculations have had success explaining nuclei in the region~\cite{Poves2014,Poves2017,Crawford2016,Miyagi2020}. Finally, (c) Two-quasi-particles plus rotor calculations (Nilsson) in the strong coupling limit~\cite{Toki1977,Kreiner78}, where the rotational energies were constrained with energy systematics of the neighbouring \isot{31,33}{Na}, and the shift of the odd-spin members was adjusted to yield a decay by two gamma rays. The proposed experimental level scheme is given in Fig.~\ref{fig:calc_summary}(d). 

\paragraph*{}
The calculations have similar features. First, all calculations have low-lying deformed $0^-, 3^-$, and $0^+$ band-heads, with associated rotational structures higher in energy. Secondly, the $B(M2)$ transition strengths are approximately correct as compared to $B(M2; 0^+ \rightarrow 2^-) = 0.23(2)$~W.u. for a $224$-keV depopulating transition. Finally, near identical single-particle occupancies are realized by all three theoretical frameworks, which are plotted in Fig.~\ref{fig:occupancies} for the $0^{+}$ and $0^-$ states and highlight the underlying $\nu f_{7/2}\rightarrow \nu d_{3/2}$ nature of the $M2$ decay. The emergence of the simplistic Nilsson scheme and its associated symmetries within the complex SDPF-U-MIX and VS-IMSRG calculations is remarkable. The spherical states, or those with dominant ``$\nu$1p-0h'' configurations, are predicted to be high in energy, $>2$~MeV (not plotted), according to the SDPF-U-MIX and VS-IMSRG predictions. For the Nilsson results, the spherical states are outside of the model space (i.e., they would require manual insertion within a two-state mixing scheme and, hence, have no predictive power).

\paragraph*{}%
In summary, a 24(2)-$\mu$s isomer in \isot{32}{Na} at the heart of the $N=20$ island of shape inversion was observed using the FDSi --- a result of excited-state spectroscopy from the first FRIB experiment. This is the only known microsecond isomer ($1{\text{ }\mu\text{s}}\leq \Thalf < 1\text{ ms}$) in the region. The odd-odd spin coupling and isomer formation provides a sensitive measure of the underlying shape degrees of freedom in a region where spherical-to-deformed shape inversion occurs. Two explanations for the isomer are given: a $6^-$ spherical shape isomer that decays by $E2$ or a $0^+$ deformed spin isomer that decays by $M2$. The present results and latest state-of-the-art calculations are most consistent with the latter, signaling that the low-lying states are dominated by deformation and that there are no clear remnants of the spherical shape after the crossing. 

\paragraph*{}
The present result provides a demonstration of the day-one scientific discovery potential of FRIB with only 1~kW of primary beam power. It also highlights the difficultly thus far in establishing reliable spectroscopic information far from stability. Over the next few years, the beam power is expected to increase in phases, corresponding to 1, 10, 100, and 400~kW. Furthermore, access to stopped and re-accelerated beams will become available. At each step, the scientific discovery potential will be expanded towards the neutron and proton drip lines and detailed spectroscopy of physics from previous steps will become possible. Starting with the increase to 10~kW, it will become feasible to resolve the spin and decay sequence of the present isomer result through $\gamma$-$\gamma$ angular correlation measurements. At the full operational power of 400~kW, the parity of the isomer could be determined through $\gamma$ polarization measurements, and the magnetic dipole moment could be measured using the time-perturbed angular distribution technique. Furthermore, mean-squared charge radii and quadrupole-moment measurements will become possible for the ground state and any millisecond isomers, through laser-assisted spectroscopy~\cite{BECOLA1,BECOLA2,RISE}. We anticipate that FRIB and the future FDS~\cite{FDS1,FDS2} at nominal operation capacity will usher in a new era of detailed spectroscopy far stability, including new discoveries at the extremes of existence.

\begin{acknowledgements}
Many thanks to B.~A.~Brown and B.~M.~Sherrill for helpful discussions. This material is based upon work supported in part by the U.S. Department of Energy, Office of Science, Office of Nuclear Physics under Contract Nos.~DE-AC02-06CH11357 (ANL), DE-AC02-98CH10946 (BNL), DE-AC02-05CH11231 (LBNL), DE-AC52-07NA27344 (LLNL), DE-SC0020451 (Michigan State), DE-SC0014448 (Mississippi State), DE-AC05-00OR22725 (ORNL), and DE-FG02-96ER40983 (UTK). The publisher acknowledges the US government license to provide public access under the DOE Public Access Plan (http://energy.gov/downloads/doe-public-access-plan). This work was also supported by the U.S. National Science Foundation under Grant Nos.~PHY-2012522 (FSU) and PHY-1848177 (CAREER) (Mississippi State). The research was also sponsored by the U.S. Department of Energy, National Nuclear Security Administration under Award No.~DE-NA0003180 (Michigan State) and the Stewardship Science Academic Alliances program through DOE Award Nos.~DE-NA0003899 (UTK) and No.~DOE-DE-NA0003906 (Michigan State), and NSF Major Research Instrumentation Program Award No.~1919735 (UTK). A.~Poves acknowledges the support of grant CEX2020-001007-S funded by MCIN/AEI/10.13039/501100011033 and PID2021-127890NB-I00. This work was supported in part by the Deutsche Forschungsgemeinschaft (DFG, German Research Foundation) --- Project-ID 279384907 --- SFB 1245.
The VS-IMSRG calculations were in part performed with an allocation of computing resources at the J\"ulich Supercomputing Center. This research was supported in part by Australian Research Council grant DP210101201. This research used resources of the Facility for Rare Isotope Beams, which is a DOE Office of Science User Facility.
\end{acknowledgements}

\bibliography{Na32_refs}

\end{document}